# THE BLACK HOLE CHALLENGE IN RANDALL-SUNDRUM II MODEL

*Nikolaos D. Pappas*[•]
Department of Physics, National KapodistrianUniversity of Athens,
Zografou University Campus, Athens, Greece

## ABSTRACT

Models postulating the existence of additional spacelike dimensions of macroscopic or even infinite size, while viewing our observable universe as merely a 3-brane living in a higher-dimensional bulk were a major breakthrough when proposed some 15 years ago. The most interesting among them both in terms of elegance of the setup and of the richness of the emerging phenomenology is the Randall-Sundrum II model where one infinite extra spacelike dimension is considered with an AdS topology, characterized by the warping effect caused by the presence of a negative cosmological constant in the bulk. A major drawback of this model is that despite numerous efforts no line element has ever been found that could describe a stable, regular, realistic black hole. Finding a smoothly behaved such solution supported by the presence of some more or less conventional fields either in the bulk and/or on the brane is the core of the black hole challenge. After a comprehensive presentation of the details of the model and the analysis of the significance and the utility of getting a specific analytic black hole solution, several (unsuccessful) analytic and numerical approaches to the problem developed over the years are presented with some discussion about their results. The chapter closes with the latest numerical results that actually consists a major advancement in the effort to address the challenge, the presentation of the most recent analytic work trying (and unfortunately failing) to build a solution assuming the existence of unconventional scalar fields and some ideas about the routes the forthcoming analytic approaches should explore.

---
[•] E-mail: npappas@cc.uoi.gr.

# INTRODUCTION

Considering higher-dimensional spacetimes is not something new in Physics. Since the formulation of the General Theory of Relativity scientists have repeatedly developed models where the key hypothesis was the existence of additional dimensions. After all, the tensorial nature of the foundations of the former and the concept of curved space-time manifold meant that the well-understood mathematical tools, developed within the context of the 4-dimensional General Relativity, could be straightforwardly generalized to include extra dimensions, both spacelike and timelike ones, and trustworthily describe such spacetimes.

On one hand, higher-dimensional models have a very appealing aspect. Since they incorporate a greater freedom for the handling of the field equations, they can be (and have been) used as a basis to develop a more unified perception of Nature. That is to address distinct phenomena in our 4-dimensional world as different, low-energy, effective projections of the same higher-dimensional entity. This quality is, quite obviously, of tremendous importance in the quest for unification and ultimately for the long-pursued "Theory of Everything". The unification effectiveness of this approach has quite successfully manifested itself in the case of the 5-dimensional Kaluza - Klein model as well as in the framework of the 11-dimensional M-theory.

On the other hand, however, several serious-and-difficult-to-address issues rise when additional dimensions come into play. For example, the existence of an extra timelike dimension poses so many and so complicated causality challenges, that this kind of models rarely get even considered. Therefore, spacelike dimensions get all the attention. The main issue then has to do with the size and the geometry of the latter. If these extra dimensions are infinitely large, like the ordinary ones, why cannot we travel along or even see them? If, on the contrary, they are so small that we could only "see" them at not-yet-reached energies, one has to deal with the puzzle of possible mechanisms that have forced them not to expand like the other three ones we know of.

In the past 35 years or so Superstring Theory, being the most promising candidate for the Theory of Everything, while demanding the existence of additional spacelike dimensions, inspired researchers to develop a multitude of higher-dimensional models in different directions. Among these models, up till recently, the too-small-to-be-observed approach was the dominant, if not the only, way scientists used to deal with the extra dimensions, the existence of which would provide them with the desired additional freedom for their models. Calabi - Yau manifolds are the most famous and characteristic representatives of this line of thinking.

However, in the turn of the last century, two novel theories were proposed, which tried to exploit the notion of branes and were grouped under the general title "brane world models". Branes (p-branes as a matter of fact) are structures that emerge in the frame of Supersting Theory and play a fundamental role in this context. They are extended objects of p spatial dimensions (strings for example are 1-branes) of whom the most important subgroup are the D-branes, on which open strings can end. Open strings describe the non-gravitational sector and their endpoints are firmly attached to branes. Closed strings, on the other hand, that describe the gravitational degrees of freedom, can propagate into the bulk. It comes quite naturally then to approach the observable Universe as a 3-brane, that is a (1+3)-hypersurface embedded in a

(1+3+n)-dimensional space-time (the bulk), with Standard Model particles and fields trapped on the brane, while gravity is free to access the bulk. This is, in a nutshell, the central idea of the brane world models (see Fig. 1).

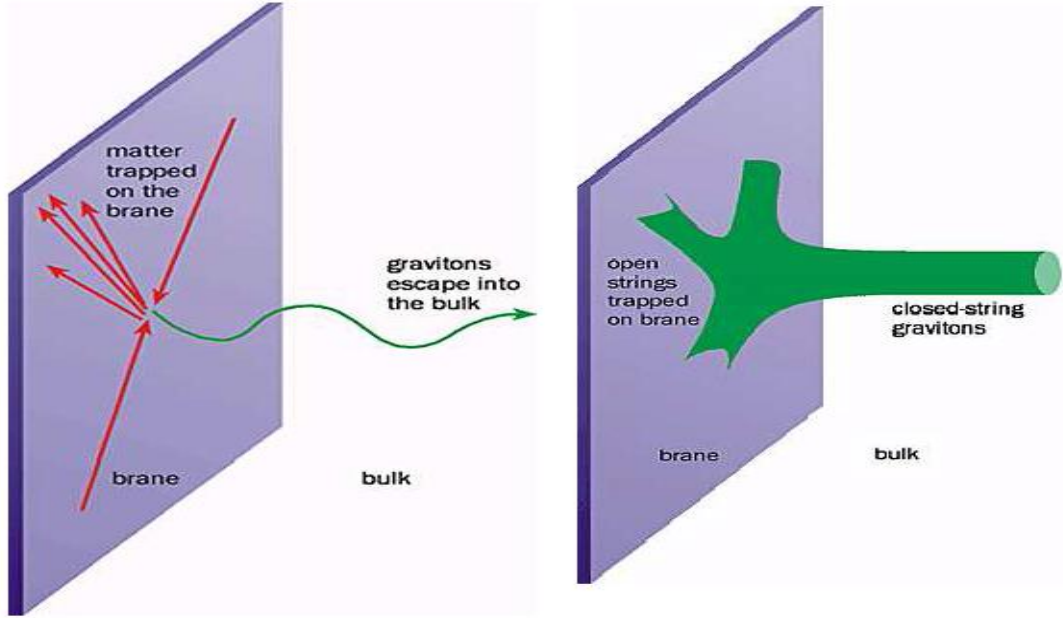

Figure 1: Diagrammatic representation of the brane world scenario, where our Universe is a 3-brane embedded in the bulk. **Left:** ordinary matter, consisting of Standard Model particle, is confined on the brane, while gravitons can propagate into the bulk. **Right**: the same fact in terms of string theory.

What was really innovative, though, was the fact that these models incorporated for the first time the idea of extra spacelike dimensions of macroscopic size. The first such model proposed back in 1998 [1] by N. Arkani-Hamed, S. Dimopoulos and G. Dvali and I. Antoniadis who postulated the existence of an arbitrary number of additional spacelike dimensions of flat topology, transverse to our 4-dimensional brane, having a size up to several μm, based on the fact that the validity of the inverse-square law for gravity has only been experimentally checked (by Cavendish-type experiments) to this limit. This model ((known as the ADD model)) drew a great deal of attention since for the first time large extra dimensions were employed and predictions that could be actually get falsified through experiment were made. Even though there are some serious conceptual problems concerning the formulation of the model, one should always acknowledge that it was the one that brought in the foreground the hypothesis that extra dimensions could be significantly different than the compactified-to-Planck-scale ones, we used to consider up till then.

Shortly after the ADD scenario was proposed another brane world model was put forward by L. Randall and R. Sundrum (RS) [2]. Actually they managed to built two, related but different models with distinct merits and problems. The trademark of the RS-models is the rather radical assumption concerning the existence of one additional spacelike dimension of infinite size transverse to our brane. The extra dimension doesn't have the trivial flat topology, though. Instead, it is characterized by the presence of a negative bulk cosmological constant $\Lambda_5$, which causes spacetime to warp and acquire an ever increasing curvature, as we move away from the brane of reference. The bulk space-time, therefore, is an anti-de-Sitter one with $\ell$, being its curvature radius related to $\Lambda_5$ as

$$\Lambda_5 = -\frac{6}{\ell^2}.$$

Another important property of the bulk is that $Z_2$-symmetry applies in it, which means that the spacetime looks exactly the same when we move away from the brane by the same distance along the extra dimension, no matter to which direction we do so. The corresponding line-element is written as

$$ds^2 = e^{-2|y|/\ell}\eta_{\mu\nu}dx^\mu dx^\nu + dy^2, \quad (1)$$

with $\eta_{\mu\nu}$ being the Minkowski metric and the $Z_2$-symmetry being realized by the presence of the factor $|y|$ in the exponent. The term $e^{-2|y|/\ell}$, usually called the "warp factor", stems from the existence of $\Lambda_5$ in the bulk and is the reason why gravity remains largely confined near the brane even though gravitons can, in principle, propagate throughout the entire (infinite) extra dimension. The brane *per se* (located at $y = 0$) has a flat Minkowski topology and we ascribe a tension to it, that represents the brane self-gravity.

From the two RS models we shall focus our attention on the one known as RS-II or single-brane RS model, as it is not only simpler and geometrically appealing, but at the same time is proven to provide a framework for AdS/CFT correspondence, while presenting a more interesting phenomenology compared to the RS-I (or two-brane) model. The main result in this context is that even though the KK modes have a continuous spectrum, the impact of the $m \neq 0$ modes on the gravitational potential is quite small, because of the warp factor. Furthermore, the massless mode (that can be seen as the massless graviton) "sees" a potential of the form

$$V(y) = \frac{15k^2}{8(k|y| + 1)^2} - \frac{3k}{2}\delta(y), \quad (2)$$

where $k \equiv 1/\ell$, that forces it to remain localized closely around the brane, thus we can speak about a bound state mode (see Fig. 2).

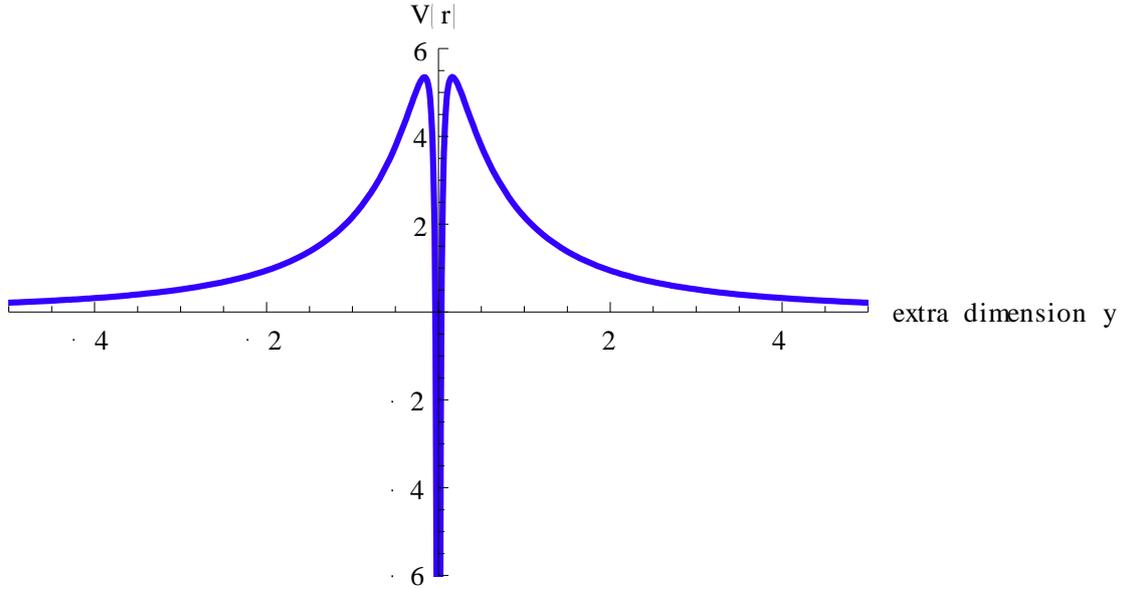

Figure 2: The "volcano" potential that causes the localization of the massless graviton in the frame of the RS-II model. The brane is located at y=0. The values on the axes mean nothing in particular as they depend on the details of the model. The important thing is the shape of the potential, which imposes this localization.

In this way the question why we don't detect the infinite extra dimension is addressed on the basis that in our low-energy experiments we deal only with the massless graviton, which is closely localized near our brane because of the warping effect of the bulk, so we have no way to directly interact with the fifth dimension in order to actually see it (remember that all other particles are strictly restricted on the brane by definition). At higher energies we would be able to study different graviton modes that would reveal the new dimension in question. The result of the aforementioned potential and the resulting confinement of the massless graviton near our brane is that the 4-dimensional gravitational potential on the latter is written as

$$V(r) \cong \frac{GM}{r}\left(1 - \frac{1}{r^2 k^2}\right) \quad (3).$$

The first term (the usual Newtonian potential) is due to the bound state mode and dominates at low energies. The correction term, that reflects the impact of all the other KK modes, becomes significant only for $r \to \ell$, so when experiments at appropriate energies would be conducted, the validity of the theory could be directly checked. I think the appeal of the model becomes easily understandable, when reflecting on the simplicity of the necessary assumptions and the specific predictions it makes. The new geometrical setup, however, requires the determination *de novo* of the predictions of the classical 4-dimensional gravity just like it was done for the gravitational potential (3). However, contrary to the ease of the calculations in the latter case, many other aspects of this phenomenology proved to be very difficult to handle. The most striking complexity associated to the RS-II background is that no stable, regular black hole solution has been found so far despite the numerous attempts towards this direction that took place in the last 15 years, as described in the following sections.

**THE IMPORTANCE OF A VIABLE BLACK HOLE SOLUTION**

The main motivation in the first place for the formulation of brane world models was the attempt to find a solution to the hierarchy problem, that is the huge discrepancy between the electroweak scale and the ordinary Planck scale, where quantum-gravitational effects arise, as observed in 4 dimensions. In such models the additional freedom that stems from the existence of extra dimensions, allows for some more radical approaches to the problem. In general, the two scales are assumed to be of the same order, while the hierarchy problem is seen as merely an artifact emerging on our brane, due to the non-trivial topology of the spacetime as a whole. Very soon, it became clear that their implications were much wider since the pioneering idea of an additional, warped and of infinite size spacelike dimension led to the reviewing of all known solutions and predictions of 4-dimensional gravity. Black-hole solutions were naturally subjected to this reviewing process as well. Interestingly enough, constructing a regular black hole metric in the context of Warped Extra Dimension Scenario proved to be a very difficult goal that is yet to be achieved. The lack of such solutions is not to be considered of secondary importance. The black hole criterion is chief among the theoretical ones that will eventually establish or refute the validity of the Warped Extra Dimension scenario as a realistic prototype of the fundamental gravitational theory, capable of addressing long-standing problems of four-dimensional physics.

There are two very important reasons why so many scientists have engaged themselves in the research for a higher-dimensional black hole solution. The first one has to do with the possibility to employ the AdS/CFT duality in order to deal with hard-to-confront issues in four dimensions starting from a simpler and classical 5-dimensional picture. It is known that the classical dynamics of the $AdS_5$ gravitational background correspond to the quantum dynamics of a 4-dimansional conformal field on the brane at linear perturbative order at least [3]. In this context the RS-II model is dual with 4-dimensional General Relativity coupled to conformal fields and, consequently, a classical 5-dimensional black hole solution in this framework is in its turn dual with a quantum-corrected black hole in four dimensions. This last connection is very interesting since it means that the (hard to calculate) quantum driven backreaction due to Hawking radiation and higher order corrections to the latter in a 4-dimensional background can be equally well described as part of the (much easier, in principle, to handle) classical dynamics of the black hole in five dimensions. Needless to say that the deeper understanding of the evaporation process of black holes (being more or less the sole region where quantum mechanics and general relativity enter on an equal footing) can provide us with valuable insights regarding the mechanisms of the long pursued theory of Quantum Gravity and this perspective makes the quest for a higher-dimensional black hole solution both important and exciting.

The other one is the prospect that naturally emerges in the case of brane world models that the creation of black holes would become significantly easier and even possible at energies accessible by current experiments. Should this prediction gets verified by reality, we would witness the birth of microscopic black holes in some accelerator collision experiment, which shortly afterwards would evaporate through the emission of Hawking radiation almost instantly. As all these phenomena would take place in front of our detectors in a well known and controlled environment, we will be able to fully detect and record this emission. Since black holes are purely gravitational objects, their behavior is determined and affected by the overall spacetime geometry. Therefore, the details of the spectrum of the emitted degrees of freedom would then provide us with solid evidence concerning the true geometry of spacetime, the existence of additional dimensions as well as the size and the topology of the latter.

And here lies the catch. In order to make the calculations necessary to determine the exact connection between the properties of the Hawking radiation spectrum and the parameters of the spacetime geometry so as to evaluate any future detector signals, one has to know the metric describing the higher-dimensional black hole in question each time. In the ADD scenario the study of black holes is straightforward since higher-dimensional (with extra dimensions being flat) versions of the Schwarzschild and Kerr solutions are known for decades (Schwarzschild-Tangherlini and Myers-Perry solutions respectively). However, in the much more interesting RS-II model (or Warped Extra Dimensions Scenario) the task of deriving a black-hole solution localized on our brane (where the gravitational collapse of matter takes place), while embedded in a curved 5-dimensional background without any spacetime singularities appearing in an uncontrollable way has proven to be unexpectedly difficult. Both analytic and numerical methods were (mostly unsuccessfully) employed to reach a solution. Even unsuccessful attempts, though, have offered valuable insights concerning the complications of the challenge and, therefore, a review of these efforts is very educative and moreover useful as basis for further research.

## ANALYTICAL APPROACHES

In the very first attempt to derive such a brane-world black-hole solution Chamblin, Hawking and Reall replaced the Minkowski part $\eta_{\mu\nu}$ of the original RS-II metric (1) with the Schwarzschild line element producing the metric [4]

$$ds^2 = e^{2A(y)}\left[-\left(1-\frac{2M}{r}\right)dt^2 + \left(1-\frac{2M}{r}\right)^{-1}dr^2 + r^2(d\theta^2 + \sin^2\theta d\varphi^2)\right] + dy^2$$

where the function $e^{2A(y)}$ is the generalized warping factor that can be reduced to the RS model one for $A(y) = -k|y|$, with k being the curvature radius of the AdS bulk spacetime. This metric succeeds in satisfying the corresponding 5-dimensional Einstein field equations. This should not be surprising since both Schwarzschild and Minkowski metrics are vacuum solutions of the field equations. Nevertheless, the aforementioned line element fails to describe a regular, localized on the brane black hole since it actually encompasses a linear singularity that extends throughout the infinite fifth dimension. The latter is clearly revealed when calculating the value of the invariant gravitational quantity $R^{MNP\Sigma}R_{MNP\Sigma}$. Then one finds

$$R^{MNRS}R_{MNRS} = \frac{48e^{-4A(y)}M^2}{r^6} + \dots \qquad (4)$$

It is obvious that for any warp function whose value reduces away from the brane (as in the RS-II case) the above quantity diverges as y goes to infinity. Even worse, equation (4) indicates the existence of a singularity at $r = 0$ for every slice of the 5-dimensional AdS spacetime related to a constant value of the y parameter. Therefore, the line element in question actually describes a black string rather than a black hole which ``escapes'' from the brane and extends up to the infinite boundaries of the fifth dimension contrary to the original motivation of the authors that proposed it. On top of that, shortly after it was shown that the black string is unstable because of the well-known from String Theory Gregory-Laflamme mechanism [5][6].

Looking more carefully at the form of the metric one can safely infer that the emergence of the black string is related to the factorized nature of the former, which means that its 4-dimensional part *per se* has no dependence on the fifth dimension coordinate (apart of course from the fact that it is multiplied with the y-dependent warp factor). It is quite natural then to assume that the restriction of the extended singularity near the brane and the reestablishment of the spacetime smoothness at a relatively short distance away from the brane could be realized through the use of a non-factorized metric, where the 4-dimensional part, observed on the brane, has an explicit dependence on the additional dimension. The choice of the right metric to do the job is not an easy task though. Earlier studies have shown that non-factorized metrics, characterized by the existence of a horizon in their 4-dimensional section, lead to spacetimes where additional singularities emerge other than the expected one at the black hole centre [7][8].

However, the following modified 4-dimensional Vaidya-type metric embedded in a 5-dimensional spacetime with a warped extra dimension seemed that it could lead to a satisfying and viable solution

$$ds^2 = e^{2A(y)}\left[-\left(1-\frac{2m(u,y)}{r}\right)du^2 + 2\varepsilon du dr + r^2(d\theta^2 + sin^2\theta d\varphi^2)\right] + dy^2 \quad (5)$$

First of all, being analytic in four dimensions the metric (5) is free of unexpected singularities. In addition to this the fact that the mass is a function of the extra dimension provides us, at least in principle, the opportunity to construct a modified, perturbed Schwarzschild-type solution in the context of which the singularity remains well-behaved and localized near the brane. Indeed, a mass function that decays faster than the square of the warp factor is capable of eliminating the singular term in the expression (4) of the curvature invariant quantity $R^{MNP\Sigma}R_{MNP\Sigma}$ within an acceptably short distance away from the brane. Despite the merits of the assumed metric, it was impossible to formulate a suitable, well-defined and functional modified version of the RS-II model that would produce a black hole solution with the desired properties. The reason is that the 4-dimensional on-brane projection of the metric (5) is no longer a vacuum solution therefore a non-trivial mass-energy distribution is necessary in order for the full 5-dimensional metric to satisfy the corresponding higher-dimensional field equations in the bulk. The functional form of the sought energy-momentum tensor was determined in [8] and was shown to satisfy all energy conditions on the brane. Furthermore, the distribution of the mass-energy related to this tensor along the extra dimension was found to have the shape of a shell which envelops the brane and thus is able to restrain the spacetime singularity of the black hole near it. Unfortunately, no self-consistent and acceptable field theory was possible to found that could give a physical explanation of this energy-momentum tensor.

The idea of bulk black holes interacting with or intersecting branes and so as to find the black hole features through the study of this interaction was considered in [9]-[12] and much more recently in [13]. The line element used by Creek, Gregory, Kanti and Mistry was the following [9]

$$ds^2 = -U(r)\,dt^2 + U(r)^{-1}dr^2 + r^2(d\chi^2 + sin^2\chi d\Omega_2^2) \quad (6)$$

with function $U(r)$ being equal to

$$U(r) = 1 + k^2 r^2 - \frac{\mu}{r^2}.$$

The analysis was based on the use of suitably modified Israel junction conditions

$$[K_{\mu\nu} - Kh_{\mu\nu}] = \kappa_5 T_{\mu\nu}$$

(with $K_{\mu\nu}$ the extrinsic curvature, $h_{\mu\nu}$ the on-brane projected metric and $T_{\mu\nu}$ the brane energy-momentum tensor, which was postulated - and hoped - to be that of a perfect fluid). From the junction conditions a set of differential equations was produced, where the trajectory of the brane $\chi(t,r)$, the brane energy density $\rho(t,r)$ and the brane equation of state $\omega(t,r)$ were the unknown parameters. Nevertheless, no parameter combination was able to produce a black hole solution with the desired properties.

A different approach was offered in [14] by Shiromizu, Maeda and Sasaki whose perspective was to incorporate the effects on the brane of the 5-dimensional Weyl tensor $C_{\mu\alpha\nu\beta}$ (see also [15]). The idea is that an observer on the brane is going to see only the part of the (generally unknown) Weyl tensor, that is projected on the brane, called the Weyl term $\epsilon_{\mu\nu}$. Then the corresponding on-bane Einstein filed equations for the observer would be

$$G_{\mu\nu}^{(4)} = 8\pi G_4 T_{\mu\nu} + \kappa_5^4 \pi_{\mu\nu} + \epsilon_{\mu\nu}$$

where the tensor $\pi_{\mu\nu}$ is quadratic with respect to the energy-momentum tensor $T_{\mu\nu}$ and consequently can safely be regarded as ignorable at the low-energy limit, where all on-brane observations are expected to take place. Furthermore, based on a series of assumptions concerning the possibility of decomposing the Weyl term into two independent parts, the relation between these parts and the asymptotic behavior of $\epsilon_{\mu\nu}$ the authors managed to produce a brane black hole solution known as the tidal Reissner-Nordstrom solution (even though no electric charge is present). However, the tidal charge appearing in the solution stems from the existence of the on-brane mass since the latter is the source of the bulk Weyl field. So here lies a cyclic and ill-understood mechanism where the on-brane mass generates a gravitational field that gets reflected back on the brane through the higher-dimensional bulk. Interesting as it may be, the overall picture remains nonetheless obscure. In addition, there are still open questions regarding the applicability of the method in the case of large black holes, the emergence of "wormholes" and the exact expression of the Weyl term.

Apart from the aforementioned four basic approaches, there are also several other papers where issues related to the black hole challenge get analytically treated like [16]-[21].

**NUMERICAL APPROACHES**

Since all efforts to find a closed-form analytic black hole solution failed in decisively addressing the challenge, numerical calculations were rather predictably the next major approach to the problem in order to provide evidence about the existence of such solutions, hints about the black hole properties and perhaps reveal the interaction between spacetime's overall geometry and the black hole behavior. Indeed, small brane-world black holes were shown to exist via numerical analysis possessing all the desired properties[1] [22]-[24] (and even in this case there were objections regarding the methods followed [25]). This kind of solutions, however, was possible to construct strictly when assuming that the black hole size is smaller than the characteristic curvature scale $\ell$ of the AdS spacetime. In this case the black hole is so small that "sees" all dimensions at an equal footing without really "feeling" the warping effect of the bulk, so that it can be approximated by a 5-dimensional Schwarzschild solution. When considering larger black holes with a horizon even the size of $\ell$ no solution could be reached, let alone the case of realistic black holes. Nature, nevertheless, seems to work in much the opposite way: small black holes may have been indeed formed in the primordial universe but none has ever been observed. Large black holes, on the other hand, with masses a million times that of the sun and macroscopic size horizons, are today believed to inhabit the center of almost every galaxy. The General Theory of Relativity allows for the analytic determination of all black-hole solutions in 4 dimensions. If the fundamental theory of gravity is indeed higher-dimensional, with its geometrical set-up being similar to the Warped Extra Dimensions Scenario, then, in principle, both small and large regular black-hole solutions should exist, thus numerical calculations should not indicate otherwise.

---

[1] That is to have a horizon at a distance from the singularity, their metric functions and derivatives to be finite (except of course at the singularity), while $AdS_5$ geometry should be recovered at asymptotic infinity.

An interesting argument, related to the existence or not of static black hole solutions in the framework of warped spacetimes, comes from the AdS/CFT correspondence point of view [26]-[28]. The general idea is that in the 4-dimensional CFT picture (which is dual to the 5-dimensional AdS picture we engaged our study so far) a black hole co-exists with a large number of conformal fields. Larger number of degrees of freedom that can be emitted means that the Hawking radiation of the black hole should be significantly enhanced. Because of the augmented magnitude of the radiation the back-reaction on the metric of the black hole mass decrease can no longer be considered negligible, thus there is no ground for a static black hole to exist. Should this argument be valid, then static localized black holes of size larger than $\ell$ (where the AdS/CFT duality holds) may not exist at all, in accordance to the results of the aforementioned numerical calculations.

In any case numerical analyses performed so far have reached non – conclusive and often contradictory results regarding realistic black hole solutions on the brane leading to arguments in favor [29]-[32] as well as against [33]-[38] their existence, only to confirm the profound difficulty of constructing such a solution already indicated by the failures of the analytic approaches mentioned earlier.

**LATEST NEWS FROM THE FRONT**

Significant progress was made in 2011 by Figueras and Wiseman [39] who managed to develop a numerical code capable of describing both large and small stable black holes within the framework of the RS-II geometry. In addition, soon after Abdolrahimi, Cattoen, Page and Yaghoobpour-Tari [40] following a different method (based on the $AdS_5/CFT_4$ duality as in the previous case) also managed to numerically produce large black hole solutions in the same background. Their findings renewed the interest of the scientific community on the subject since it became evident that the elusive-up-to-now black hole solutions do exist, despite all the doubts and disappointments accumulated over the years.

In 2013 another attempt to analytically address the challenge was launched by Kanti, Pappas and Zuleta [41] who went back to an earlier idea employing once again the modified 5-dimensional Vaidya-type metric (5). Allowing the mass to be a function of both the fifth and the time coordinate, this metric provides a reasonable ansatz for a perturbed Schwarzschild background on the brane, ideal for investigating both the localization of the black hole singularity as well as the existence of a static solution. In the previous use of (5) in [8] only ordinary theories of scalar or gauge fields were accounted, which failed to provide a viable solution. In the new approach the authors allowed themselves a greater freedom regarding the nature of the scalar fields considered. The idea was to assume that the black hole mass function has an exponential form capable of canceling the singular term in the value of the invariant curvature quantity $R^{MNP\Sigma}R_{MNP\Sigma}$ appearing in eq. (4) (which is responsible for the infinitely long black string result discussed in the introduction), while reducing to the Schwarzschild picture on the brane. Then taking advantage of the freedom incorporated in the model (warp factor to be y-dependent but otherwise of arbitrary form, no fine-tuning between brane and bulk parameters) we thought that we could find a specific model that would satisfy the 5-dimensional field equations. The viable bulk solution that would emerge could finally be used to determine the brane content thought the junction conditions. First minimally coupled to gravity but otherwise described by a general Lagrangian scalar fields were studied. A detailed analysis was performed in the cases of a single scalar with a non-canonical kinetic term, its Lagrangian being

$$L_{sc} = \sum_{n=1} f_n(\varphi)(\partial^M \varphi \partial_M \varphi)^n + V(\varphi),$$

two interacting scalars with Lagrangian

$$L_{sc} = f^{(1)}(\varphi, \chi)\partial^M \varphi \partial_M \varphi + f^{(2)}(\varphi, \chi)\partial^M \chi \partial_M \chi + V(\varphi, \chi),$$

two interacting scalars with general kinetic terms with Lagrangian

$$L_{sc} = \sum_{n=1} f_n^{(1)}(\varphi, \chi)(\partial^M \varphi \partial_M \varphi)^n + \sum_{n=1} f_n^{(2)}(\varphi, \chi)(\partial^M \chi \partial_M \chi)^n + V(\varphi, \chi)$$

and finally two interacting scalar fields with mixed kinetic terms described by

$$L_{sc} = f^{(1)}(\varphi, \chi)\partial^M \varphi \partial_M \varphi + f^{(2)}(\varphi, \chi)\partial^M \chi \partial_M \chi + f^{(3)}(\varphi, \chi)\partial^M \varphi \partial_M \chi + V(\varphi, \chi).$$

Unfortunately, all field configurations failed to satisfy the corresponding 5-dimensional field equations. Furthermore, as the same analysis can be straightforwardly generalized to allow for more general kinetic and mixing terms, the general result was that models with one or two no minimally-coupled-to-gravity scalars simply cannot do the job.

Next the case of non-minimally coupled scalars was studied. The following action was considered

$$S = \int d^4x\, dy \sqrt{-g}\left[\frac{f(\varphi)}{2\kappa_5^2}R - \frac{1}{2}(\nabla\varphi)^2 - V(\varphi) - \Lambda_5\right]$$

where both the exact form of $f(\varphi)$ and $\varphi$ are arbitrary to avoid any unreasonable restrictions on the field configurations. After checking the implications of a coupling function being a power law, a polynomial and an exponential function of the field and proving that all these are dead-end choices, a more general analysis was employed for a $f(\varphi)$ of completely arbitrary form. Finally, a no-go argument was formulated stating that a model of a non-minimally coupled scalar field is altogether inconsistent with the desired black hole mass behavior, as described earlier.

Two major conclusions were derived from this work. The first was that the localization of the black hole appears to demand the synergetic action from both the brane and the bulk parameters. The second rather interesting outcome was that even in the case where the black hole mass was postulated to be time-independent, the additional fields required to support the model had to be dynamic. A static black hole configuration was not excluded by the calculations, but nevertheless was shown to be quite hard to build. As we feel that the potential of this method is yet to be exhausted, we have already started a new research program to explore a) whether more delicate mass function variations could be the answer and b) the possibility that a reasonable scalar field configuration could support a viable black hole solution that is not strictly Schwarzschild on the brane, but rather Schwarzschild-like.

Furthermore, a specific analytic calculation based on the $AdS_5/CFT_4$ duality remains to be done following the steps of the respective numerical results mentioned in the previous section in order to exploit the insights offered by them. In this case the starting point is to consider an exact Schwarzschild metric at a brane located at the infinite boundary of the AdS spacetime. This gravitational background on the brane, rewritten in a more general way, can get expanded along the bulk to produce a Randall-Sundrum brane at a finite proper distance whose induced metric is a perturbed Schwarzschild metric and thus describes a black hole. Then by solving the 5-dimensional field equations one would, in principle, completely determine all model parameters. Obviously, doing all that analytically is far from trivial but in the light of the certainty (thanks to the latest numerical calculations) that realistic black hole solutions are "out there" things just might be a bit easier and some optimism is to be allowed.

**CONCLUSION**

Up to now, the Warped Extra Dimensions Scenario, although one of the most popular ever suggested in theoretical physics, has failed to pass the black-hole test.

Finding a regular black hole solution in this framework is a major challenge not only because it has been proven a not-at-all trivial task but also because the existence of such a solution would enhance the possibilities of this scenario to be a realistic description of the actual spacetime geometry. Furthermore, due to the AdS/CFT duality a classical 5-dimensional black hole metric would allow for the determination of the corresponding quantum-corrected 4-dimensional metric, thus providing us with new, deeper insights about the correlations between quantum mechanics and general relativity. Besides, if strong gravity effects emerge indeed in the few TeV energy regime, possessing an explicit black hole solution is the key factor in order to evaluate the related detector signals to find solid evidence about the existence of extra spacelike dimensions, their number and their geometry.

The prize is too significant to be ignored and that is why since the formulation of the RS models until nowadays the scientific community never ceased to try to discover a viable black hole solution. The quest is still active and open for new ideas, methods and techniques. Hopefully, in the next few years solid results would be reached to elucidate the situation and decisively answer the long lasting questions concerning the existence and the exact form of black holes in the context of the warped extra dimension scenario with all the far reaching implications that should accompany such a discovery.